\documentclass[epj,doublespace]{svjour}
\usepackage{graphicx}
\usepackage{epstopdf}
\usepackage{amssymb}
\usepackage{bm}
\usepackage{amsmath}
\usepackage{nicefrac}
 \usepackage[table]{xcolor}
\usepackage{siunitx}
\usepackage{tikz}
\usepackage{enumitem}
\usepackage{comment}
\usepackage{array,multirow}
\usepackage[normalem]{ulem}
\usepackage{setspace}
\usepackage{dcolumn}

 \usepackage{hyperref}
\usepackage{soul}
\begin{document}
\title{
  On the separability of microscopic optical model potentials and
  emerging bell-shape Perey-Buck nonlocality
}
\author{
  H. F. Arellano\inst{1} 
  \and 
  G. Blanchon\inst{2,3}
}                     
%
%

\institute{
  Department of Physics - FCFM, University of Chile,
  Av. Blanco Encalada 2008, Santiago, RM 8370449, Chile
 \and
  CEA,DAM,DIF, F-91297 Arpajon, France
  \and
  Universit\'e Paris-Saclay, CEA, Laboratoire Mati\`ere en 
  Conditions Extr\^emes,Bruy\`eres-le-Ch\^atel, France
          }
%


\date{Received: date / Revised version: date}
%
\abstract{
After nearly sixty years since its introduction,
the phenomenological bell-shape
Perey-Buck spatial nonlocality in the optical model
potential for nucleon-nucleus scattering has remained 
unaccounted for from a microscopic standpoint.
In this article we provide a quantitative account for such nonlocality
considering fully nonlocal optical potentials in momentum space.
The framework is based on a momentum-space \textit{in-medium} 
folding model,
where infinite nuclear matter $g$ matrices in
Brueckner-Hartree-Fock approximation are folded to the
target one-body mixed density. 
The study is based on chiral next-to-next-to-next-to-leading order (N3LO) 
as well as Argonne $v_{18}$ nucleon-nucleon bare interaction models.
Applications focus on $^{40}$Ca($p,p$) scattering at beam energies 
in the range 11--200~MeV, 
resulting in the identification of a separable structure 
of the momentum-space optical potential of a form
we coin as \textit{JvH},
with a nonlocality form factor as one of its terms.
The resulting nonlocaliy form factor features a bell-shape with
nonlocality range $\beta$ between 0.86 and 0.89~fm, 
for both proton and neutron beams at energies below 65~MeV.
An analytic toy model is introduced to elucidate the underlying
mechanism for the nonlocality in the optical model, 
providing an estimate of its range based on the Fermi motion
of the target nucleons and the long-range part
of the \textit{NN} interaction.
}
\authorrunning{Arellano-Blanchon}
\titlerunning{On the separability of microscopic optical...}
\maketitle
 \sloppy
%




\section{Introduction}
The fermionic nature of nucleons together with many-body
correlations constitute two essential aspects for conceiving nonlocality 
in the coupling between nucleonic probes and nuclei.
This feature was early investigated in the late 50s with the introduction 
of phenomenological optical model potentials with Gaussian 
nonlocality~\cite{Frahn1956,Frahn1957,Wyatt1960}.
Subsequent studies by Perey and Buck in the context of neutron 
scattering resulted in the 
construction of an energy-independent nonlocal optical 
potential~\cite{Perey1962},
with reasonable account of scattering data 
for aluminum to lead targets at energies in the range $0.4-\!24$~MeV.
The merit of this construction is that the parameters of the potential
were obtained from fits to differential cross section data 
for $^{208}$Pb$(n,n)$ at two energies only, namely 7 and 14.5~MeV.
The range of nonlocality of the assumed Gaussian form factor
was found to be around 0.85~fm,
quantity broadly accepted nowadays but still unaccounted for 
from a microscopic standpoint.

Some estimates for the range of nonlocality in optical potentials
for nucleon scattering have been 
reported~\cite{Ripka1963,Giannini1976,Jeukenne1976},
in addition to investigations aimed to identifying its 
causes~\cite{Rawitscher1985,Fraser2008}.
Additionally, several efforts have been pursued in order to
broaden the extent of Perey-Buck-like 
parametrizations~\cite{Tian2015,Lovell2017,Jaghoub2018}.
These works are strongly motivated by the recognized role of 
nonlocal effects in processes such as $(p,d)$ transfer and capture 
reactions~\cite{Ross2015,Bailey2017,Tian2018},
in addition to dispersive optical 
models~\cite{Dickhoff2010,Waldecker2011,Mahzoon2014}.

In this paper we give a quantitative account for the nonlocality 
in the optical potential for nucleon-nucleus scattering,
in addition to its shape. 
This is achieved after a close scrutiny of 
the momentum-space structure of 
microscopic optical potential based on chiral 
N3LO~\cite{Entem2003}
as well as Argonne $v_{18}$ (AV18)~\cite{Wiringa1995}
bare nucleon-nucleon (\textit{NN}) interactions.
We show that, to lowest order, the optical potential
factorizes in such a way that its nonlocality form factors
can be isolated and calculated microscopically.
This feature allows parameter-free accounts for the 
nonlocality shape and range, 
in addition to information on radial form factors
as well as volume integrals.

%
%
%
%
%

\section{Framework}
The nonrelativistic optical potential in momentum space 
for nucleon elastic scattering off spin-zero nucleus 
can be cast as~\cite{Ray1992}
\begin{equation}
  \label{ukk}
  \tilde {\cal U}({\bm k'},{\bm k};E) =
  \tilde {\cal U}_{c}({\bm k'},{\bm k};E) +
  i{\bm\sigma}\!\cdot\hat{\bm n}\,\,
  \tilde {\cal U}_{so}({\bm k'},{\bm k};E)\;,
\end{equation}
with $E$ the center-of-mass (c.m.) energy,
${\bm\sigma}$ twice the spin of the projectile and
$\hat{\bm n}$ 
given by
$\hat{\bm n}\!=\!{\bm k'\!\times\!\bm k}/{|\bm k'\times\bm k|}$.
Here $\tilde {\cal U}_{c}$ and $\tilde {\cal U}_{so}$ 
denote the central and spin-orbit components of the potential.
To evaluate the potential we follow 
Refs. \cite{Arellano1995,Arellano2007a,Aguayo2008},
with an infinite nuclear-matter model to represent the 
\textit{in-medium} effective \textit{NN} interaction.
Additionally, the use of the Slater approximation for the one-body 
mixed density yields~\cite{Arellano1995}
\begin{align}
  \label{folding}
  \tilde {\cal U}_{\nu}({\bm k'},{\bm k};E) =&
  4\pi\!\!\sum_{\alpha=p,n}
  \int_0^{\infty}\!\!\! z^2dz\,\rho_\alpha(z) j_0(qz) 
  \nonumber \\
  \times\int\! d{\bm P}& S_z(P)
  \left\langle \textstyle{\frac{{\bm k}'-{\bm P}}{2}}|
    g_{\bar K}^{\nu\alpha} (E+\bar\epsilon)
         |\textstyle{\frac{{\bm k}-{\bm P}}{2}}\right\rangle_{\!\cal A}\;,
\end{align}
where ${\cal A}$ denotes antisymmetrization,
$\nu$ the nucleonic probe and $\rho_\alpha(z)$ 
the nuclear density of species $\alpha$ at coordinate $z$.
Additionally, $g^{\nu\alpha}$ is the 
density-dependent Brueckner-Hartree-Fock (BHF) $g$ matrix, 
evaluated at the isoscalar density $\rho(z)$,
coupling nucleon $\nu$ with target nucleon $\alpha$.
In the above $S_z$ is given by
\begin{equation}
  \label{Sz}
  S_z(P)=\frac{3}{4\pi k_z^3}\, \Theta(k_z\!-\! P),
\end{equation}
setting bounds for the off-shell sampling of the $g$ matrix at the
radial coordinate $z$.
The local Fermi momentum $k_z$ depends on
the isoscalar density $\rho(z)$ at the radial coordinate $z$ through
\begin{equation}
  \label{kz}
   k_z= \left [ \frac{3\pi^2\rho(z)}{2}\right ]^{1/3}.
\end{equation}
Applications of this parameter-free model have been reported in Refs.
\cite{Arellano1995,Arellano2002,Aguayo2008}, 
allowing reasonable descriptions of elastic scattering at nucleon 
energies from tens of MeV up to about 1~GeV.

The calculations of microscopic optical model potentials we
carry out are based on proton and neutron densities from Gogny's 
D1S Hartree-Fock calculations~\cite{Decharge1980}.
Additionally, fully off-shell \textit{NN} $g$ matrices at fifteen 
equally spaced Fermi momenta $k_F$ 
are calculated in momentum space
solving the BHF equation for symmetric nuclear matter 
\begin{equation}
  \label{bhf}
  \hat g(\omega) = \hat v + 
  \hat v \,
  \frac{\hat Q}{\omega+i\eta - \hat h_1 - \hat h_2}\,
  \hat g(\omega)\;,
\end{equation}
with $\omega$ the starting energy,
$\hat v$ the bare \textit{NN} interaction,
$\hat Q$ the Pauli blocking operator suppressing nucleon
propagation below the Fermi energy, and
$\hat h_{1,2}$ nucleons single-particle energies.
To solve this integral equation we apply
techniques introduced in Ref.~\cite{Arellano2015}.
We stress here that self-consistent $g$ matrix calculations
do take into account Cooper eigenvalues in $^1$S$_0$ and
$^3\textrm{S}_1\!-
\!^3\textrm{D}_1$~\cite{Arellano2015,Arellano2016,Isaule2016}.
Scattering observables for the resulting nonlocal
optical potentials are obtained using 
{\small SWANLOP} package~\cite{Arellano2021} together with
{\small SIDES} package~\cite{Blanchon2020} for double-checks,
both treating nonlocalities without approximations.

\begin{figure}[ht]
  \includegraphics[width=0.95\linewidth] {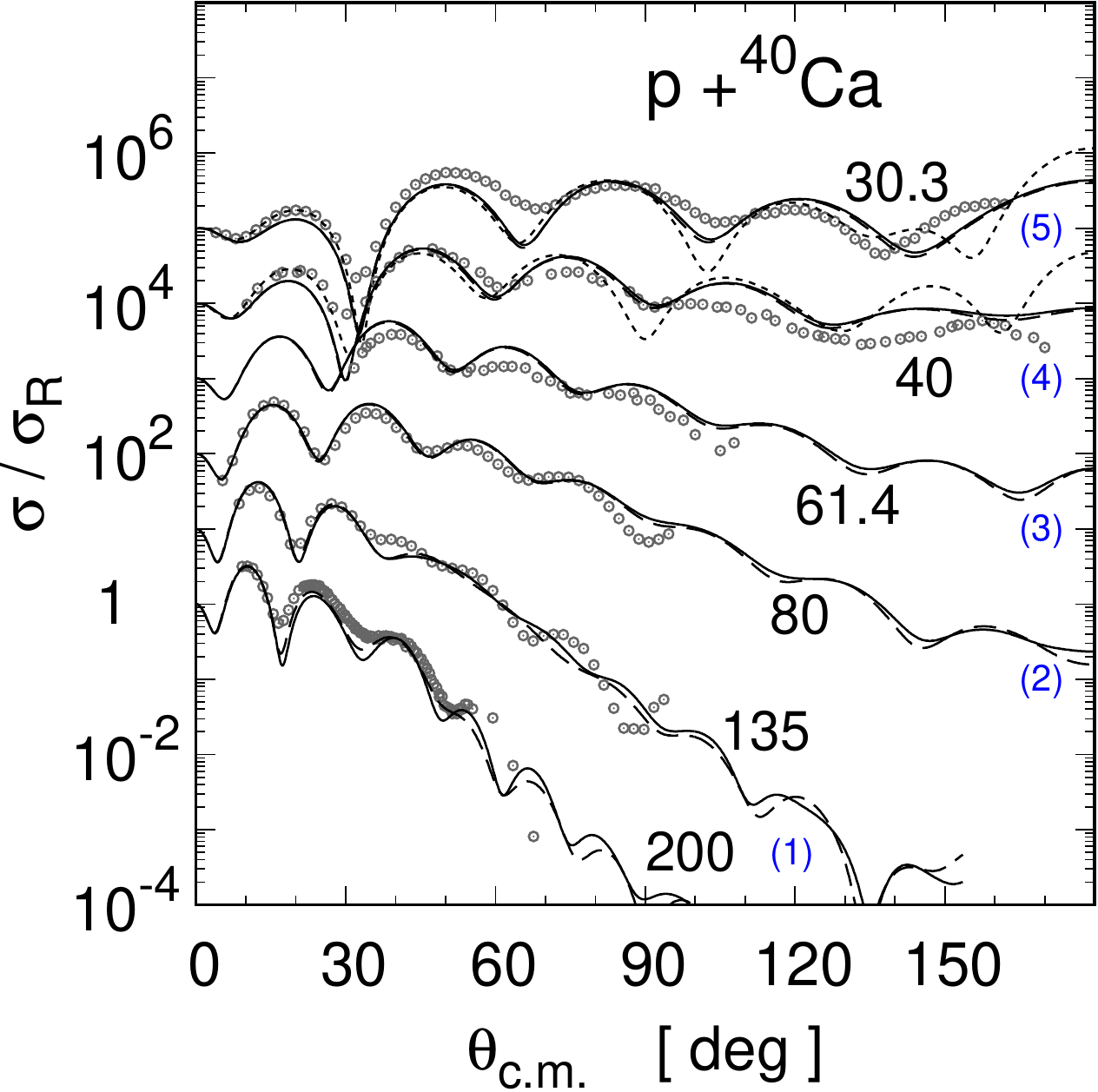}
\medskip
\caption{{\protect\small
\label{dsdw}
Ratio-to-Rutherford differential cross sections
   as function of the c.m. scattering 
  angle for $^{40}$Ca$(p,p)$ scattering. 
  Solid and long-dashed curves correspond to momentum-space $g$-matrix 
  folding potentials using chiral N3LO and AV18 bare interactions, 
  respectively. 
  Short-dashed curves correspond to Perey-Buck model using 
  TPM parametrization.
  Labels indicate beam energy in MeV units. 
  Numbers in parentheses indicates the power of ten in upshift.
  The data are from 
  Refs.~\cite{Ridley1964,Fricke1967,Fulmer1969,Nadasen1981,Hutcheon1988}
        }
        }
\end{figure}

The consistency of the momentum-space approach adopted here
is illustrated 
in Fig.~\ref{dsdw}, where we plot the calculated
differential cross section for proton elastic scattering off
$^{40}$Ca at beam energies in the set 
${\cal E}_6\!=\!\{30.3, 40, 61.4, 80, 135, 200\}$,
in MeV units.
Solid and dashed curves denote results based on N3LO and 
AV18 bare interactions, respectively.
Nonlocal Perey-Buck-based parametrization by 
Tian \emph{et al.}~\cite{Tian2015} (TPM) 
{at 30.3 and 40~MeV}
are shown with short-dashed curves.
The data are from
Refs.~\cite{Ridley1964,Fricke1967,Fulmer1969,Nadasen1981,Hutcheon1988}.
We observe that results based on N3LO and AV18
bare potentials yield comparable descriptions of the data, 
except at 200~MeV where N3LO shows deeper diffraction minima.
In the case of TPM parametrization at 40~MeV and below, 
the calculated 
ratio-to-Rutherford differential cross sections $\sigma(\theta)/\sigma_R$
appears more diffractive than 
those from microscopic approaches. 
Overall, we can state that the parameter-free  microscopic
approach is able to grasp leading features in the scattering process
over a broad energy range.

\subsection{Separability}
\begin{figure}[ht]
  \begin{center}
\includegraphics[width=0.75\linewidth] {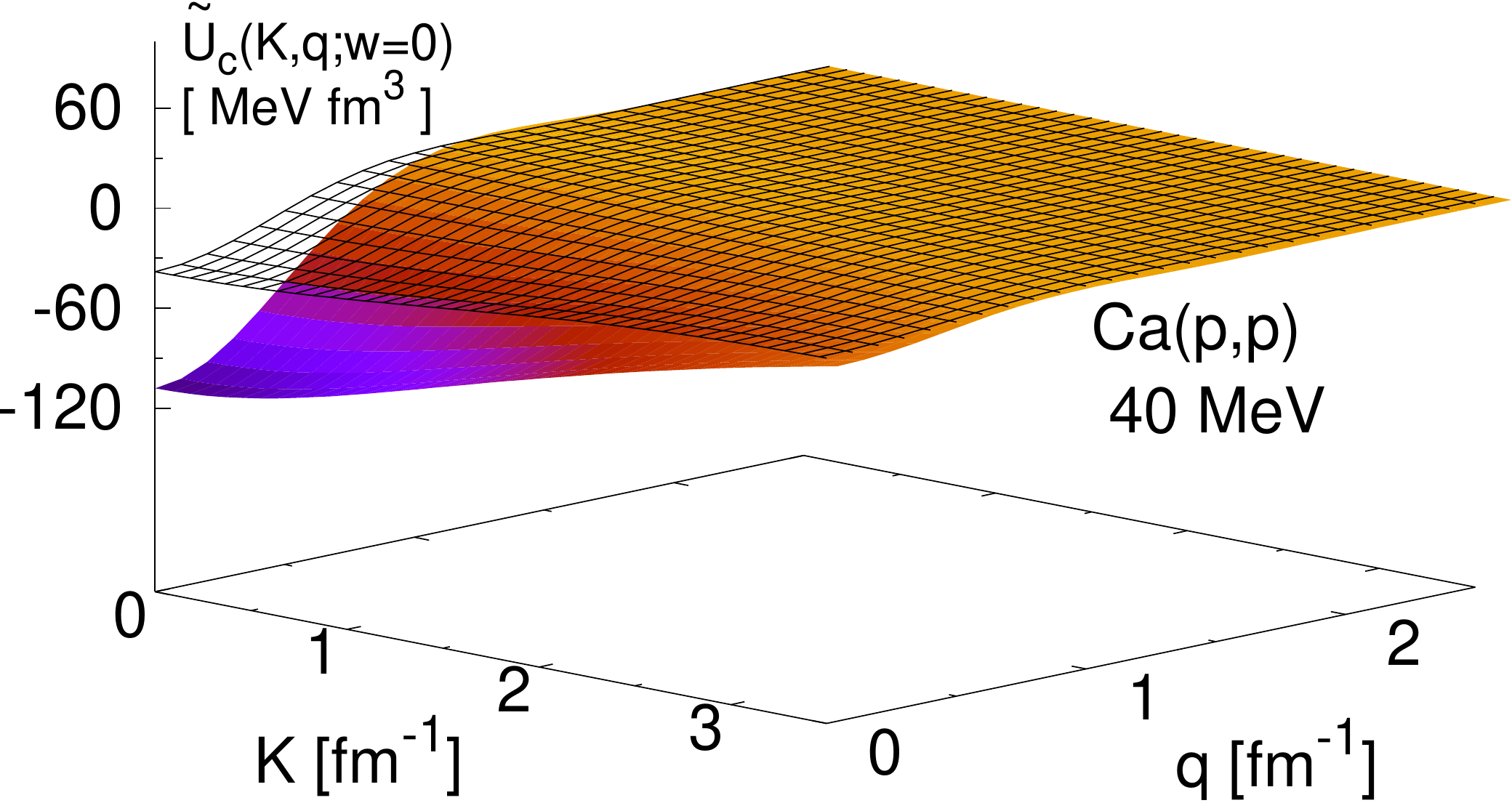}
\end{center}
\medskip
\caption{{\protect\small
\label{Uc}
  Surface plots in the $Kq$ plane of 
  the central optical potential $\tilde U_c(K,q;w=0)$, 
  for $^{40}$Ca$(p,p)$ scattering at 40~MeV.
  The colored surface represents its real part and the black mesh
  its imaginary component.
  Results from momentum-space $g$-matrix folding potentials 
  based on AV18 bare \textit{NN} potential.
        }
        }
\end{figure}

To examine the momentum-space structure of the potential we find 
advantageous to re-express it in terms of 
momentum transfer ${\bm q}$, mean momentum ${\bm K}$ 
and the cosine of their relative angle:
\begin{subequations}
  \begin{align}
    {\bm q}&={\bm k} - {\bm k'} \;;\\
    {\bm K}&=\textstyle{\frac12} \, ( {\bm k} + {\bm k'})\;;\\
          w&=\hat{\bm K}\!\cdot\! \hat{\bm q}\;.
  \end{align}
\end{subequations}
Accordingly, we define 
$\tilde U({\bm K},{\bm q})\!=\!\tilde {\cal U}({\bm k'},{\bm k};E)$,
so that
\begin{equation}
  \label{notation}
{\tilde U}({\bm K},{\bm q})
  = {\tilde U}_{c}(K,q;w) +
  i {\bm\sigma}\cdot({\bm K}\times{\bm q})\, {\tilde U}_{so}(K,q;w)\;,
\end{equation}
leaving the dependence on $E$ implicit.
In Fig.~\ref{Uc} we show a surface plot of $\tilde U_c(K,q,w)$
in the $Kq$ plane for the case of Ca($p,p$) scattering at 40~MeV. 
For this plot we have chosen ${\bm K}\perp{\bm q}$, namely $w=0$.
The real part of the potential is shown with colored surface,
whereas its imaginary part is shown with a black mesh.
We observe that the potential exhibits a smooth behavior
in both, $K$ and $q$ momenta, decreasing for increasing momenta.

\begin{figure}[ht]
\includegraphics[width=0.95\linewidth] {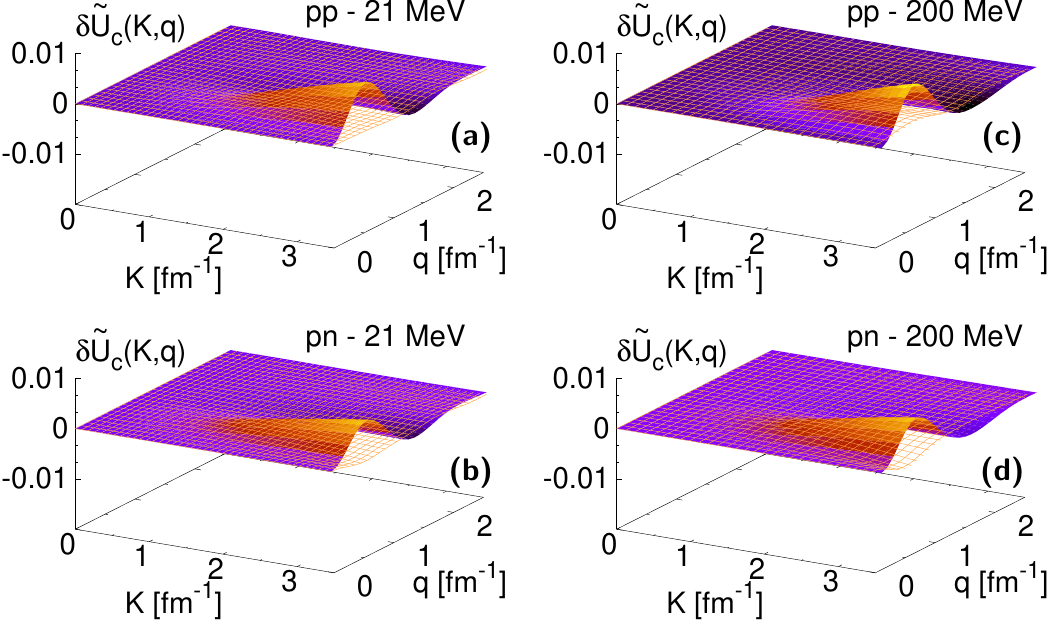}
\medskip
\caption{{\protect\small
\label{deltaUc}
  Surface plots in the $Kq$ plane for the real and imaginary components of 
  $\delta\tilde U_c(K,q)$, for $^{40}$Ca$(p,p)$ scattering
  at 21(a,b) and 200~MeV (c,d).
  Results from momentum-space $g$-matrix folding potentials 
  based on AV18 bare \textit{NN} potential.
        }
        }
\end{figure}

We have investigated the $w$-dependence of $\tilde U_c$ and $\tilde U_{so}$
for microscopic potentials at all energies in ${\cal E}_6$,
including lower energies, finding it very weak.
To illustrate the weak angular dependence we have calculated the
normalized difference
\begin{equation}
  \label{udiff}
  \delta \tilde U_c(K,q)=\frac{1}{\tilde U_c(0,\!0;\!0)}
  \left [
\tilde U_c(K,q;w_1)-\tilde U_c(K,q;w_2)
  \right ]\;,
\end{equation}
considering $w_1\!=\!0$, and $w_2\!=\!1$.
Under this definition $\delta\tilde U_c$ becomes dimensionless and
equal to unity for $K\!=\!q\!=\!0$.
We have examined $\delta\tilde U_c$ separately 
for the coupling of the projectile
(protons) to target protons ($pp$) and target neutrons ($pn$).
In Fig.~\ref{deltaUc} we show surface plots of 
$\delta\tilde U_c(K,q)$ as a function of $K$ and $q$, at
beam energies of 21 (a,b) and 200~MeV (c,d).
The real parts are shown with colored surfaces whereas 
their imaginary parts are shown with orange meshes.
As observed, the difference is negligible for 
$K\!\lesssim\!1.5$~fm$^{-1}$, exhibiting some structure
below $\sim\!1$\% at $q$ below $1$~fm$^{-1}$.
Similar features are observed for the spin-orbit term.

The above results justify to ignore (to lowest order)
the $w$-dependence in the optical potential, assigning
\begin{subequations}
\begin{align}
  \label{pw1}
    {\tilde U}_{c}(K,q;w) &\to {\tilde U}_{c}(K,q)\;, \\
  \label{pw2}
  |{\bm K}\times{\bm q}|\, {\tilde U}_{so}(K,q;w) &\to
    Kq \sqrt{1-w^2} \,{\tilde U}_{so}(K,q) \,.
  \end{align}
\end{subequations}
As a way to remove the $w$ dependence we use $w\!=\!0$, 
choice that ensures symmetric relative momenta, i.e. $k\!=\!k'$.

To disentangle the structure of $\tilde U_{c}(K,q)$ and 
$\tilde U_{so}(K,q)$ 
in the $Kq$ plane, let us define $W_{c}\!=\!\tilde U_{c}(0,0)$ and 
  \begin{subequations}
    \begin{align}
\tilde V_{c}(K,q)&=\frac{\tilde U_{c}(K,q)}{\tilde U_{c}(K,0)}; \label{vv} \\
      \tilde H_{c}(K)&=\!\frac{\tilde U_{c}(K,0)}{\tilde U_{c}(0,0)}; \label{hh} 
    \end{align}
  \end{subequations}
so that
  \begin{equation}
  \label{separation1}
  \tilde U_{c}(K,q) = W_{c}\;\tilde V_{c}(K,q) \;\tilde H_{c}(K)\;.
  \end{equation}
An analogous construction is applied to the spin-orbit term, resulting
in
  \begin{equation}
  \label{separation2}
  \tilde U_{so}(K,q) = W_{so}\;\tilde V_{so}(K,q) \;\tilde H_{so}(K)\;.
  \end{equation}
Note that 
$\tilde V_c(K,q)$, $\tilde H_c(K)$,
$\tilde V_{so}(K,q)$ and $\tilde H_{so}(K)$
are dimensionless quantities satisfying 
$\tilde V_{c}(K,0)\!=\tilde V_{so}(K,0)\!=\!1$, and 
$\tilde H_c(0)\!=\tilde H_{so}(0)\!=\!1$.
It is also worth stating that,
in connection to Perey-Buck notation~\cite{Perey1962},
$(K,q)$ momenta are the conjugate variables of their $(s,R)$ coordinates.

\begin{figure}[ht]
\includegraphics[width=0.95\linewidth] {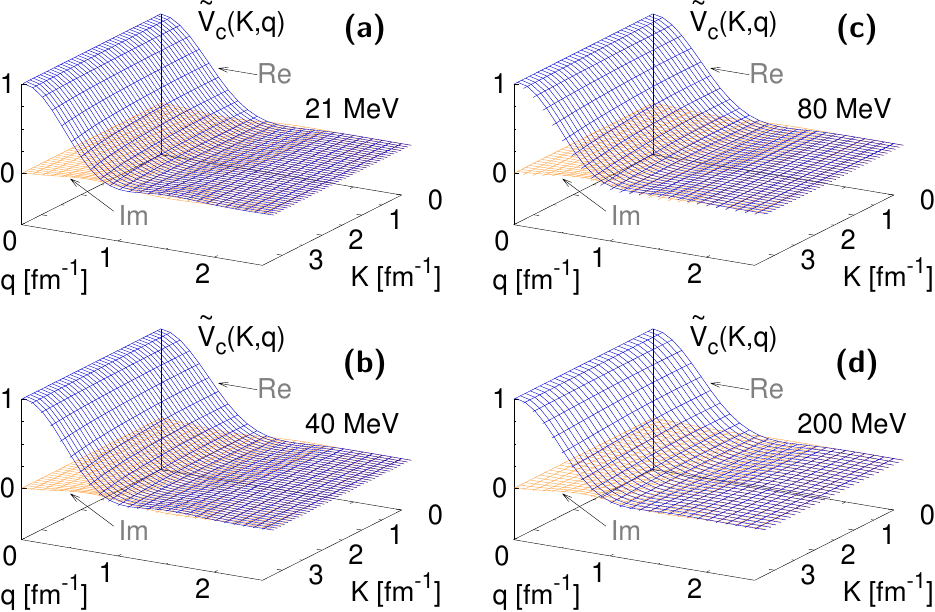}
\medskip
\caption{{\protect\small
\label{v_Kq}
  Surface plots in the $Kq$ plane for the real and imaginary components of 
  $\tilde V_c(K,q)$, for $^{40}$Ca$(p,p)$ scattering
  at 21(a), 40(b), 80(c) and 200(d) MeV.
  Results from momentum-space $g$-matrix folding potentials 
  based on AV18 bare \textit{NN} potential.
        }
        }
\end{figure}

We now proceed to examine separately 
the three terms on the right-hand-side of 
Eqs.~\eqref{separation1} and \eqref{separation2}.
We have investigated the momentum dependence of 
$\tilde V_{c}(K,q)$ and $\tilde V_{so}(K,q)$ 
at all energies in ${\cal E}_6$.
We found that their dependence on $K$ is remarkably weak.
This feature is illustrated in Fig.~\ref{v_Kq}, 
where we show surface plots of $\textrm{Re}\,\tilde V_c(K,q)$ 
and $\textrm{Im}\,\tilde V_c(K,q)$, in the $Kq$ plane.
The energies considered here are 
21 (a), 40 (b), 80 (c) and 200 (d) MeV.
Note the close resemblance of all four surfaces, 
despite the broad energy range they span,
exhibiting small imaginary parts in all cases.
These features are also observed in the spin-orbit term $\tilde V_{so}$.

\begin{figure}[ht]
\includegraphics[width=0.95\linewidth] {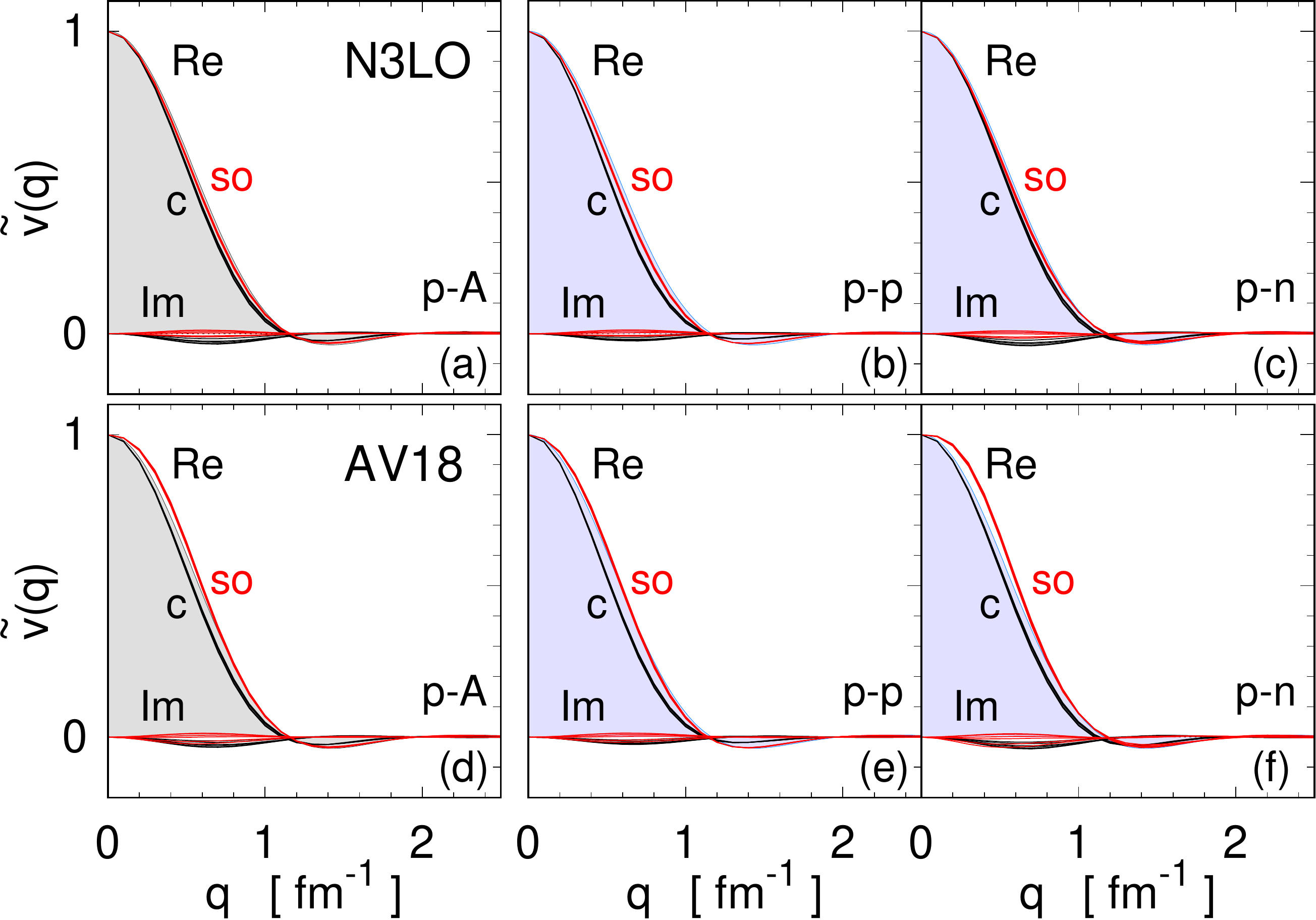}
\medskip
\caption{{\protect\small
\label{vq}
  Radial form factor $\tilde v_c(q)$ for $^{40}$Ca$(p,p)$ 
  scattering based on N3LO and AV18 bare $NN$ potentials.
  Panels (a,d), (b,e) and (c,f) show results for 
  $p\,\text{-}A$, $p\,\text{-}p$ and $p\,\text{-}n$ couplings, 
  respectively.
        }
        }
\end{figure}

The potential $U_{pA}$ for proton-nucleus scattering 
can be decomposed as 
\begin{equation}
  \label{UpA}
  U_{pA}=U_{pp}+U_{pn}\;,
\end{equation}
where the subscripts $pp$ and $pn$ denote coupling of the incident
proton with target protons and neutrons, respectively.
To each of these three terms we investigate the factorization
\eqref{separation1}, labeling them as 
$p\,\text{-}A$, $p\,\text{-}p$ and $p\,\text{-}n$, respectively.
Considering the weak dependence of $\tilde V_c(K,q)$ and 
$\tilde V_{so}(K,q)$ on $K$, in Fig.~\ref{vq} we plot 
the radial form factors
$\tilde v_{c}(q)\!\equiv\!\tilde V_c(0,q)$, and
$\tilde v_{so}(q)\!\equiv\!\tilde V_{so}(0,q)$,
as functions of $q$ for all energies in ${\cal E}_6$.
Upper (Lower) panels correspond to potentials based on N3LO (AV18)
bare \textit{NN} interactions.
Central (c) and spin-orbit (so) components are plotted 
with black and red curves, respectively.
Panels (a,d) correspond to $p\,\text{-}A$ coupling;
       (b,e)            to $p\,\text{-}p$ coupling; and
       (c,f)            to $p\,\text{-}n$ coupling.
The shaded areas denote 
Fourier transforms of the isoscalar, 
proton and neutron densities for $^{40}$Ca used in these calculations.
The similarity of all six curves,
particularly in their real parts,
point to a weak energy dependence of both central and spin-orbit terms. 
The imaginary component shows more differences,
but remaining weaker relative to their real parts.
When comparing N3LO- and AV18-based potentials, 
differences are noticed in the real spin-orbit (red curves).
Such differences suggest shorter range in $\textrm{Re}\,\tilde v_{so}$ 
for the AV18- relative to N3LO-based interactions.

\begin{figure}[ht]
\includegraphics[width=0.95\linewidth] {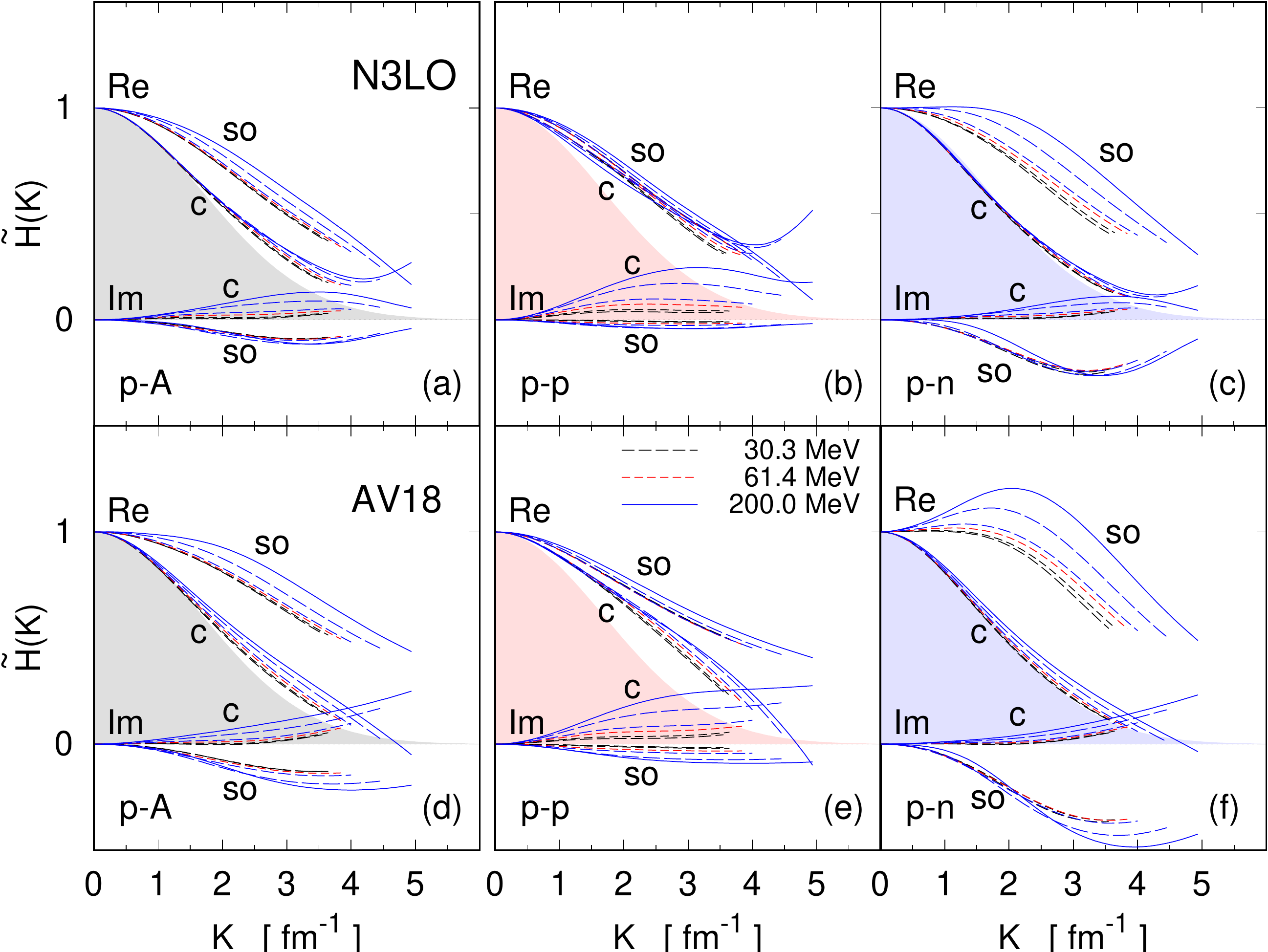}
\medskip
\caption{{\protect\small
\label{hk}
Bell-shape nonlocality form factor $\tilde H(K)$ from momentum-space
$g$-matrix folding potentials in energy set ${\cal E}_6$.
Results for central (c) and spin-orbit (so),
  based on N3LO and AV18 bare \textit{NN} potentials.
Panels (a,d), (b,e) and (c,f) show form factors for
  $p\,\text{-}A$, $p\,\text{-}p$ and $p\,\text{-}n$ couplings, 
  respectively.
The shaded areas   denote
Perey-Buck nonlocality form factor
with $\beta\!=\!0.84$~fm.
        }
        }
\end{figure}

Let us now focus on $\tilde H_{c}(K)$ and $\tilde H_{so}(K)$, 
both directly related to the nonlocality of the potential. 
In the Perey-Buck model the nonlocal form factor in coordinate space 
is defined as
\begin{equation}
    \label{hpb}
      H_{P\!B}(s) = \frac{1}{{\pi}^{3/2} \beta^3} e^{-s^2/\beta^2}\;,
\end{equation}
with $s$ the difference between prior and post relative coordinates
and $\beta$ the range of nonlocality. 
In momentum space, 
\begin{equation}
  \label{HPBK}
  \tilde H_{P\!B}(K)=e^{-\beta^2K^2/4}\;.
\end{equation}
As done for $\tilde v(q)$ in Fig.~\ref{vq}, we analyze
separately the $p\,\text{-}A$, $p\,\text{-}p$ and $p\,\text{-}n$ terms.
In Fig.~\ref{hk} we plot $\tilde H(K)$ 
calculated from momentum-space folding potentials at all 
energies in ${\cal E}_6$.
Panels and labels are the same as in Fig.~\ref{vq}.
Solid black and blue curves correspond to proton beam energy of
30.3 and 200~MeV, respectively.
Dashed red curves represent results for 61.4~MeV.
The shaded areas represent
Perey-Buck form factor with $\beta\!=\!0.84$~fm.

From Fig.~\ref{hk} we observe that $\textrm{Re}\,\tilde H_c(K)$ for 
N3LO- and AV18-based potentials exhibit similar patterns at all energies,
all of them featuring a bell shape.
In particular, both $p\,\text{-}A$ and $p\,\text{-}n$ appear quite similar
to Perey-Buck's form factor $\tilde H_{P\!B}$, 
in contrast to $p\,$-$p$ where weaker curvatures
takes place at the origin. 
This feature implies that for $p\,\text{-}p$ coupling, 
the associated nonlocality range is smaller than that
for $\tilde H_{P\!B}$.
With regard to the spin-orbit form factor, we notice that
$\textrm{Re}\,\tilde H_{so}(K)$ is quite disperse for 
$p\,$-$A$ and $p\,$-$n$, particularly above
61.4~MeV (dashed red curves).
Furthermore, curvatures of $\tilde H_{so}$ are consistently
smaller than those for $\tilde H_{c}$, pointing to smaller range
of nonlocality relative to the central part.
We also note that $\textrm{Re}\,\tilde H_{so}(K)$ follow a similar
trend, although their curvatures near $K\!=\!0$ appear weaker.  

In Table~\ref{tab:betap} we tabulate the calculated $\beta$ 
as function of the proton energy from Gaussian fits of
$\textrm{Re}\,\tilde H(K)\!=\!\exp(-\beta^2 K^2/4)$,
over the range $0\!\leq\!K\!\leq\!1$~fm$^{-1}$.
To illustrate the trend of these results at lower energies
  we include 11.42 and 21~MeV.
We observe that $\beta_{pA}$ in the central channel
varies 3\% at most at energies below 80~MeV, 
for both N3LO- and AV18-based microscopic potentials.
We also note that $\beta_{pn}\!-\!\beta_{pp}\!\approx\!0.22$~fm, 
in all cases below 100~MeV,
indicating that the coupling of protons to target neutrons is 
slightly more nonlocal than the coupling to protons. 
A similar trend is observed in calculations made for 
neutron scattering.
The above values for the range of nonlocality
are consistent with PB reported value
$\beta_n\!=\! 0.85$~fm, 
and TPM values
$\beta_n\!=\! 0.90$~fm, and
$\beta_p\!=\! 0.88$~fm.
\newcolumntype{g}{>{\columncolor{gray!25}}c}
\begin{table*}[ht] 
  \begin{tabular}{|c|c| g c c | g c c |}
\hline
  \hspace{3mm} & & \multicolumn{3}{c|}{N3LO} & \multicolumn{3}{c|}{AV18}\\
       \hline
  &Energy & 
    $\hspace{6pt}\beta_{pA}$\hspace{6pt} & 
    $\hspace{6pt}\beta_{pp}$\hspace{6pt} & 
    $\hspace{6pt}\beta_{pn}$\hspace{6pt} & 
    $\hspace{6pt}\beta_{pA}$\hspace{6pt} & 
    $\hspace{6pt}\beta_{pp}$\hspace{6pt} & 
    $\hspace{6pt}\beta_{pn}$\hspace{6pt} \\
  &(MeV) & (fm) & (fm) & (fm) & (fm) & (fm) & (fm) \\
  \hline
  \parbox[t]{2mm}{\multirow{8}{*}{\rotatebox[origin=c]{90}{Central}}}
  &11.42 & 0.89 & 0.72  & 0.94  & 0.89  & 0.73  & 0.95 \\
  &21.0  & 0.88 & 0.72  & 0.94  & 0.89  & 0.72  & 0.94 \\
  &30.3  & 0.88 & 0.71  & 0.93  & 0.88  & 0.71  & 0.94 \\
  &40.0  & 0.88 & 0.71  & 0.93  & 0.88  & 0.71  & 0.93 \\
  &61.4  & 0.87  & 0.72  & 0.93  & 0.86  & 0.70  & 0.92 \\
  &80.0  & 0.87  & 0.72  & 0.93  & 0.86  & 0.71  & 0.92 \\
  &135.0  & 0.87  & 0.75  & 0.92  & 0.83  & 0.71  & 0.89 \\
  &200.0  & 0.86  & 0.78  & 0.90  & 0.80  & 0.72  & 0.85 \\
  \hline
  \parbox[t]{2mm}{\multirow{8}{*}{\rotatebox[origin=c]{90}{Spin-orbit}}}
  &11.42	&0.57	&0.61 	&0.51 	&0.47 	&0.58 	&0.03 \\
  &21.0	        &0.58 	&0.61 	&0.50 	&0.46 	&0.59 	&---  \\
  &30.3	        &0.58 	&0.62 	&0.49 	&0.48 	&0.59 	&---  \\
  &40.0	        &0.58 	&0.63 	&0.48 	&0.48 	&0.60 	&---  \\
  &61.4	        &0.57 	&0.63 	&0.43 	&0.46 	&0.60 	&---  \\
  &80.0	        &0.55 	&0.62 	&0.37 	&0.37 	&0.59 	&---  \\
  &135.0	&0.48 	&0.59 	&0.13 	&0.32 	&0.56 	&---  \\
  &200.0	&0.44 	&0.56 	&---  	&0.22 	&0.51 	&---  \\
\hline
\end{tabular}
  \caption{\label{tab:betap}
  Central and spin-orbit nonlocality $\beta_p$ 
  for $p\!+\!^{40}$Ca scattering as functions
  of the energy from momentum-space folding potentials 
  based on chiral N3LO and AV18 bare interactions.}
\end{table*}

In the spin-orbit sector of Table~\ref{tab:betap}, 
entries with dashes correspond to cases with positive curvature 
for $\tilde H$ at the origin.
This feature occurs in the $p\,\text{-}n$ 
coupling using AV18 bare potential.
This peculiarity gets manifested in $\beta_{pA}$, 
resulting in a more local $p\,\text{-}A$ coupling for AV18- relative 
to N3LO-based potential.
What is remarkable from these results is that the spin-orbit
term of the optical potential also exhibits a bell-shape nonlocality, 
although such nonlocality appears smaller than in the central part
by about $0.3 -\!0.4$~fm. 
This is in contrast with Perey-Buck model,
where the spin-orbit term is local.
This result is interesting in the context of phenomenological
constructions of nonlocal potentials, suggesting a 
prescription to include nonlocality in the spin-orbit term.

  We have also investigated the range of nonlocality $\beta$ in the
  case of neutron scattering. 
  Results are summarized in Table~\ref{tab:betan}, for neutron energies
  between 10 and 65~MeV, 
  following the same convention as in Table~\ref{tab:betap}.
  As observed, $\beta_{nA}$ is very similar to $\beta_{pA}$ for
  both, central and spin-orbit components.
\newcolumntype{g}{>{\columncolor{gray!25}}c}
\begin{table*}[ht] 
\begin{tabular}{|c|c| g c c | g c c |}
\hline
  \hspace{3mm} & & \multicolumn{3}{c|}{N3LO} & \multicolumn{3}{c|}{AV18}\\
       \hline
  &Energy & 
    $\hspace{6pt}\beta_{nA}$\hspace{6pt} & 
    $\hspace{6pt}\beta_{np}$\hspace{6pt} & 
    $\hspace{6pt}\beta_{nn}$\hspace{6pt} & 
    $\hspace{6pt}\beta_{nA}$\hspace{6pt} & 
    $\hspace{6pt}\beta_{np}$\hspace{6pt} & 
    $\hspace{6pt}\beta_{nn}$\hspace{6pt} \\
  &(MeV) & (fm) & (fm) & (fm) & (fm) & (fm) & (fm) \\
  \hline
  \parbox[t]{2mm}{\multirow{5}{*}{\rotatebox[origin=c]{90}{Central}}}
  &11.9  & 0.89 & 0.94  & 0.72  & 0.90  & 0.95  & 0.73 \\
  &21.7  & 0.88 & 0.94  & 0.72  & 0.89  & 0.94  & 0.72 \\
  &30.3  & 0.88 & 0.93  & 0.71  & 0.88  & 0.94  & 0.71 \\
  &40.0  & 0.88 & 0.94  & 0.71  & 0.88  & 0.94  & 0.71 \\
  &65.0  & 0.88 & 0.93  & 0.71  & 0.86  & 0.92  & 0.70 \\
  \hline
  \parbox[t]{2mm}{\multirow{5}{*}{\rotatebox[origin=c]{90}{Spin-orbit}}}
  &11.9 &0.57 	&0.50 	&0.60 	&0.47 	&0.03 	&0.58 \\
  &21.7 &0.57 	&0.50 	&0.60 	&0.47 	&---  	&0.58 \\
  &30.3 &0.57 	&0.49 	&0.61 	&0.47 	&---  	&0.59 \\
  &40.0 &0.58 	&0.48 	&0.62 	&0.47 	&---  	&0.60 \\
  &65.0 &0.56 	&0.42 	&0.62 	&0.45 	&---  	&0.60 \\
\hline
\end{tabular}
  \caption{\label{tab:betan}
  Central and spin-orbit nonlocality $\beta_n$ 
  for $n\!+\!^{40}$Ca scattering as functions
  of the energy from momentum-space folding potentials 
  based on chiral N3LO and AV18 bare interactions.}
\end{table*}

\begin{figure}[ht]
\includegraphics[width=0.97\linewidth] {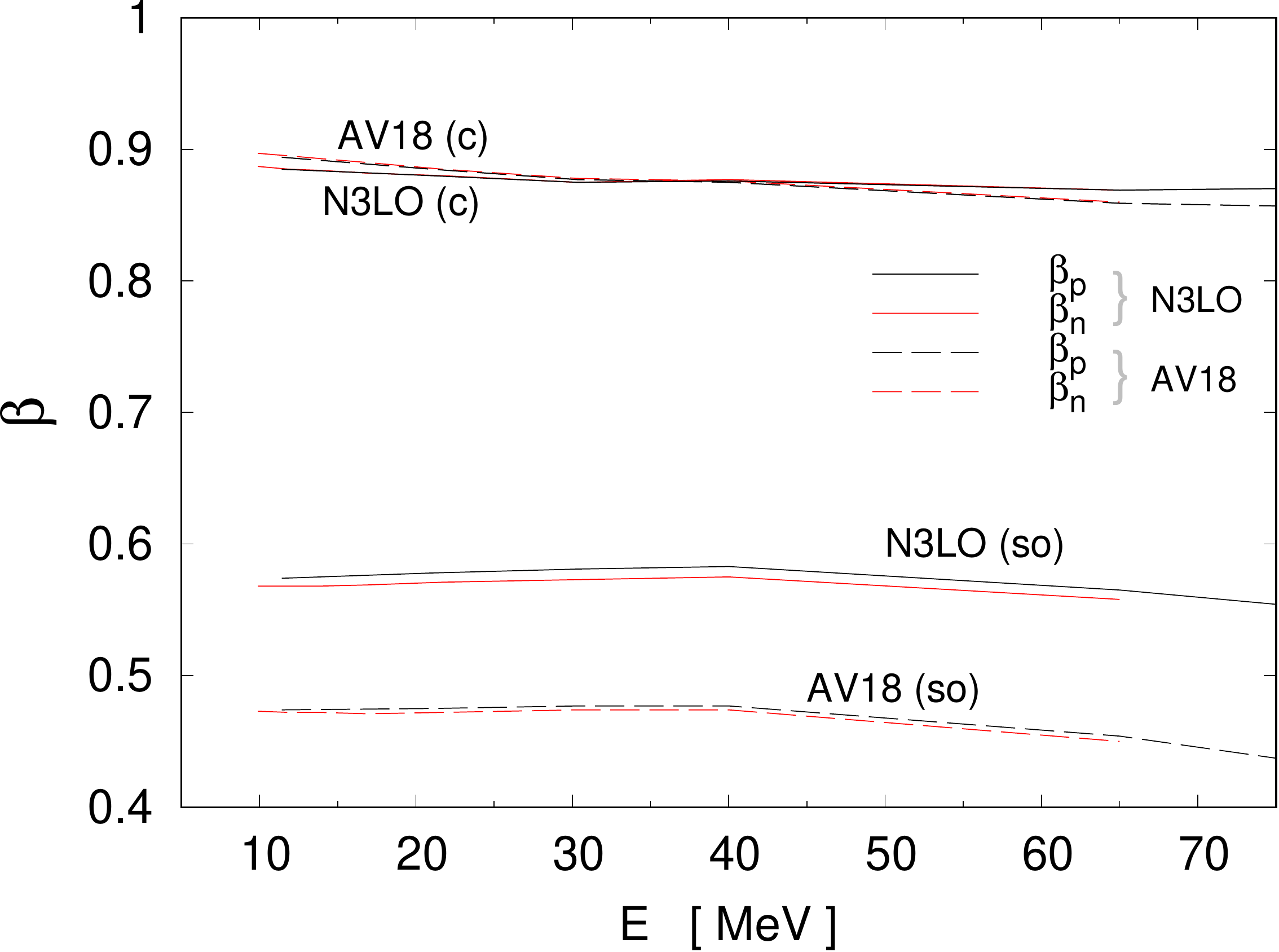}
\medskip
\caption{{\protect\small
\label{betan}
  Energy dependence of the range of nonlocality for proton ($\beta_p$) 
  and neutron ($\beta_n$) scattering off $^{40}$Ca. 
  Results for the central (c) and spin-orbit (so) parts of
  the potential, based on chiral N3LO (solid curves) and
  AV18 (dashed curves) bare interactions.
        }
        }
\end{figure}

  In Fig.~\ref{betan} we plot $\beta$ as a function of the energy 
  for proton (black curves) and neutron (red curves) scattering 
  off $^{40}$Ca, to be called $\beta_p$ and $\beta_n$, respectively. 
  Solid and dashed curves denote results based on N3LO and AV18 
  interactions, respectively.  
  We include results for the central (c) and spin-orbit (so).
  As observed, for a given \textit{NN} interaction, $\beta_p$ 
  and $\beta_n$ in the central part are quite similar to each other,
  with differences below $\sim$1\% when comparing N3LO with AV18.
  In the case of the spin-orbit term, under a given bare interaction
  the differences between $\beta_p$ and $\beta_n$ are below 1\%.
  However, the nonlocality obtained from N3LO becomes larger by 
  about 1~fm relative to that based on AV18.
  Overall, Fig.~\ref{betan} shows weak energy dependence of $\beta$ 
  in all cases.

\begin{figure}[ht]
\includegraphics[width=0.97\linewidth] {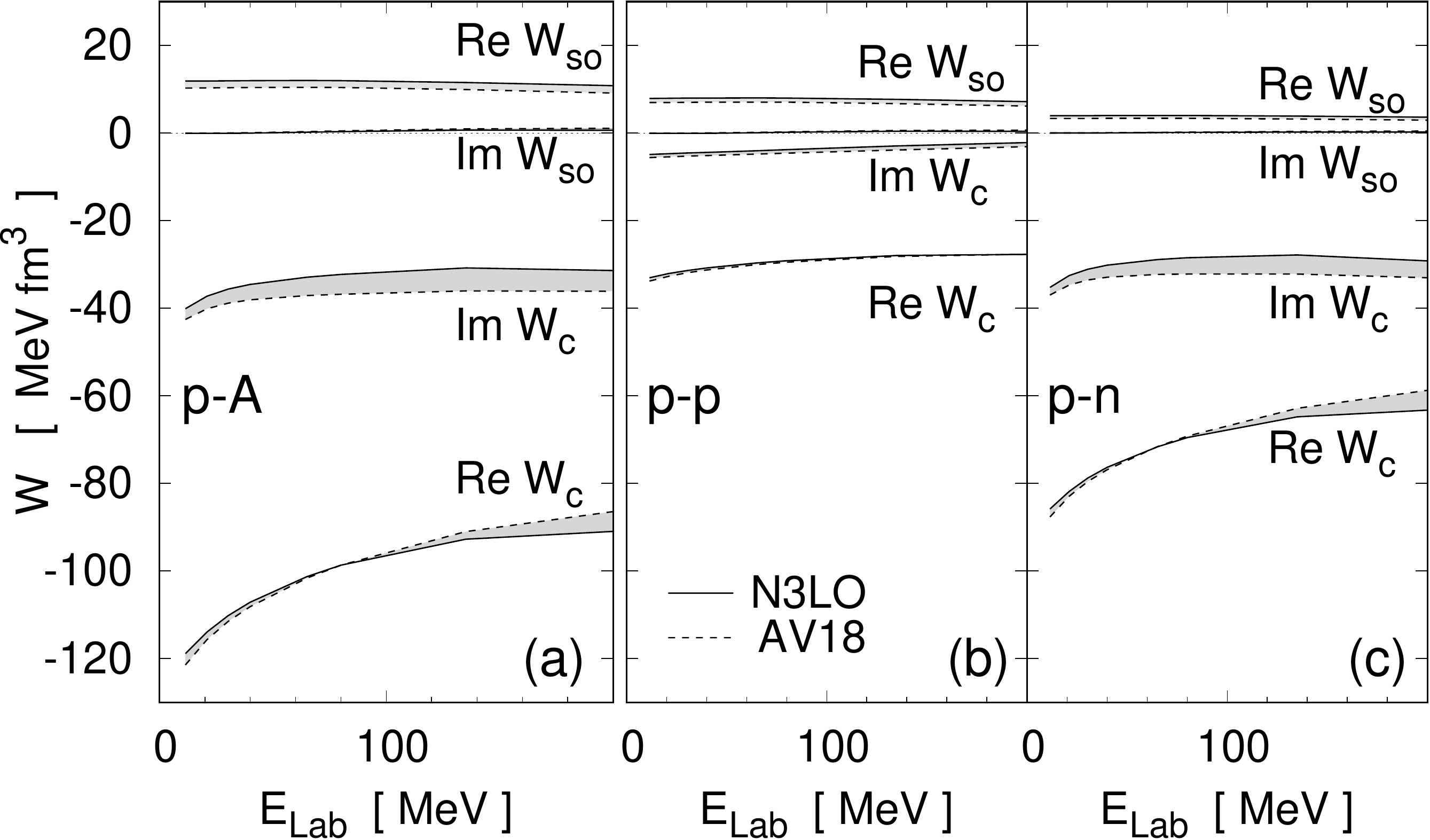}
\medskip
\caption{{\protect\small
\label{w}
 Energy dependence of the real and imaginary components of volume 
 strengths $W_c$ and $W_{so}$.
  Results for $^{40}$Ca$(p,p)$ scattering considering 
N3LO (solid curves) and AV18 (dashed curves) bare potentials.
  Panels (a), (b) and (c) show results for 
  $p\,\text{-}A$, $p\,\text{-}p$ and $p\,\text{-}n$ couplings, 
  respectively.
        }
        }
\end{figure}

Another term in factorization  \eqref{separation1} is $W$, 
corresponding to the potential for zero momenta.
This strength is directly related to the volume integral $J$ of
the potential~\cite{Arellano2021}, namely 
\begin{equation}
  \label{J}
  W=\frac{J}{(2\pi)^3}\;.
\end{equation}
In Fig.~\ref{w} we show the central ($W_c$) and
spin-orbit ($W_{so}$) strength as function of the beam energy,
where solid (dashed) curves denote results based on N3LO (AV18)
bare interaction.
We observe that the $p$-$n$ strength is stronger than its $p$-$p$
counterpart in all channels.
It also becomes evident that both $p$-$n$ and $p$-$p$ parts are nearly
energy independent, 
except for $\textrm{Re}\,W_c$ in the $p\,\text{-}n$ coupling
which changes by 12\% in the range $11\!-\!40$~MeV.
The spin-orbit strengths $W_{so}$, in turn,
are quite small relative to the central part, 
nearly vanishing for $\textrm{Im}\,W_{so}$ in $p\,\text{-}n$ coupling.

\subsection{The \textit{JvH} factorization}
In order to assess the ability of the three factors in 
Eq.~\eqref{separation1} to retain the original description
of $\sigma/\sigma_R$, we introduce the separable representation
\begin{align}
  \label{separation3}
  \tilde U(K,q) = &W_{c} \, \tilde v_{c}(q) \, \tilde H_{c}(K) \nonumber \\
  & + i\,{\bm \sigma}\cdot ( {\bm K}\times{\bm q})
  \,W_{so} \, \tilde v_{so}(q) \, \tilde H_{so}(K)\;.
\end{align}
We will refer to this decomposition as \textit{JvH} factorization,
being comprised by the strength $W\!=\!J/(2\pi)^3$, 
the normalized radial form factor $\tilde v$ and
the normalized nonlocality form factor $\tilde H$.
Implicit in this construction is that the 
$w$-dependence of $\tilde U_c$ and $\tilde U_{so}$ in Eq.~\eqref{notation},
and $K$-dependence of $\tilde V(K,q)$, are weak.

\begin{figure}[ht]
\includegraphics[width=0.95\linewidth] {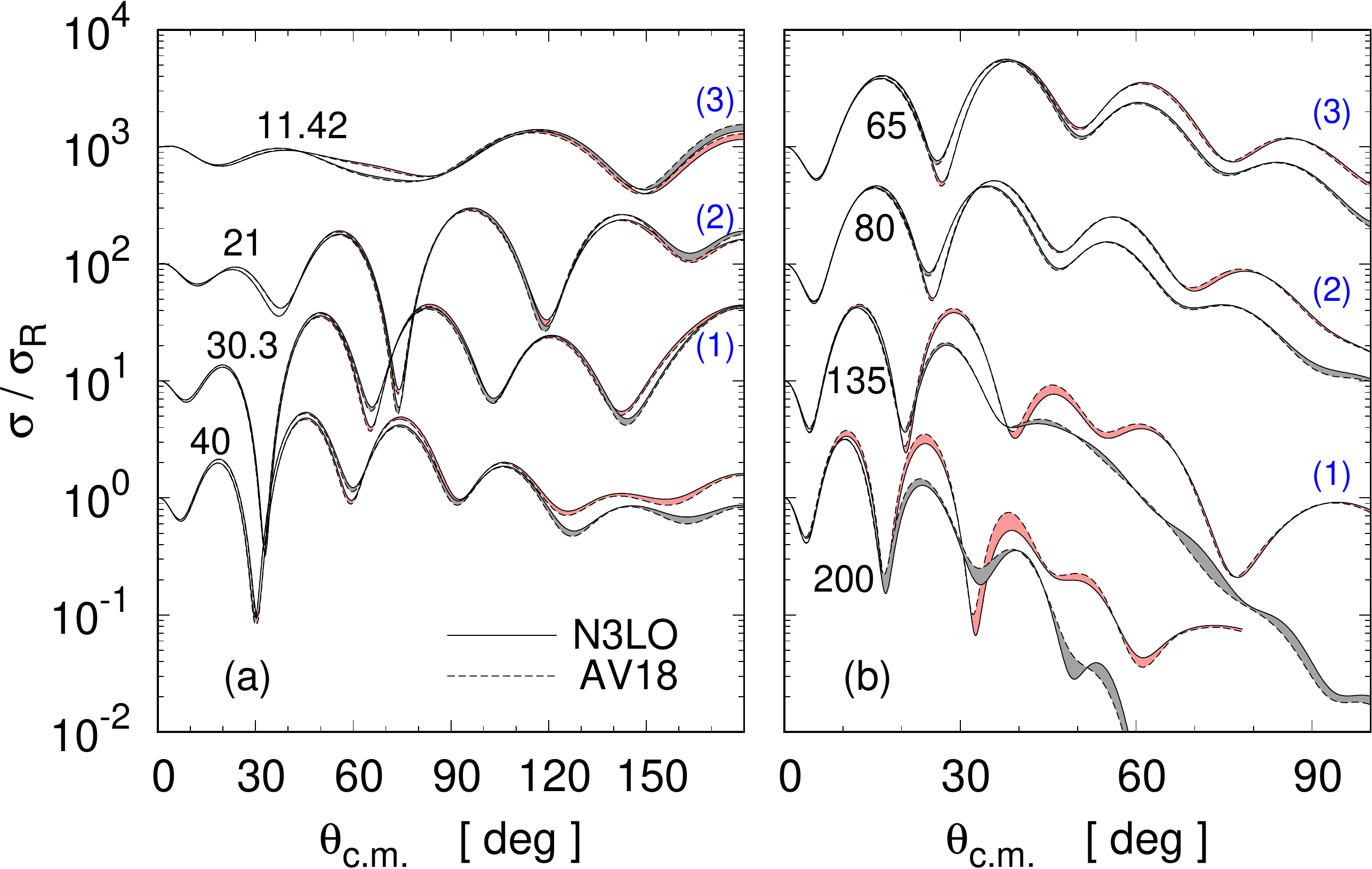}
\medskip
\caption{{\protect\small
\label{check}
Ratio-to-Rutherford differential cross sections 
as functions of $\theta_{c.m.}$
for 
$^{40}$Ca$(p,p)$ scattering 
from momentum-space folding potentials (gray shaded areas) 
and its \textit{JvH} form (red shaded areas).
  Solid (short-dashed) curves denote results using
  N3LO (AV18) interactions.
  Labels indicate beam energy in MeV units. 
  Numbers in parentheses denote power of ten of upshift.
        }
        }
\end{figure}

In Fig.~\ref{check} we show results for the 
ratio-to-Rutherford differential 
cross sections for $^{40}$Ca$(p,p)$ 
scattering from momentum-space folding potentials (gray shaded) and 
  corresponding \textit{JvH} form (red shaded).
Solid and short-dashed curves denote N3LO- and AV18-based results, 
respectively.
Results are shown as functions of the c.m. scattering angle
$\theta_{c.m.}$.
In these plots we include applications at 11.4 and 21~MeV.
As observed in panel (a) at energies up to 40~MeV, 
close agreement takes place between the original potential 
and its \textit{JvH} representation.
This is noted by comparing gray- vs red-shaded areas, although
discrepancies appear at 40~MeV above 120~deg.
In panel (b) for the higher energies, however,
differences become evident at angles above the second minimum,
roughly around 25~deg.

An important element for the separability of the optical potential 
is its weak dependence on $w\!=\!\hat K\cdot\hat q$.
This is in line with a recent study of 
Burrows \textit{et al.}~\cite{Burrows2018} on features of the 
nonlocal one-body density 
[$\rho({q}, \mathcal{K})$ in their notation].
The authors also find that the largest contribution to the nonlocal
density comes from $s$ waves, with an associated nonlocality
of about 2~fm.
In actual calculations of the optical potential, these features
in the one-body mixed density need to be convoluted with the 
effective interaction. 
Its implications on the nonlocality of the optical potential
is illustrated in the following section with the introduction
of a toy model.

\subsection{Toy model}
An interesting outcome from this study is that the 
microscopic momentum-space optical model potential summarized 
by Eq.~\eqref{folding} accounts consistently for the 
range of nonlocality in PB and TPM models obtained phenomenologically.
We now devise a toy model aiming to understand
the role of physical elements responsible of the range of nonlocality
in the optical model.
The parameter-free approach used here involves the 
folding of an antisymmetrized Fermi-motion-averaged 
effective interaction.
With these elements in mind let us
consider a uniform nuclear density over a sphere of radius $R$ with 
bulk Fermi momentum $k_z\!\to\!\bm \bar k_F$.

In the Slater approximation the one-body mixed density, 
$\rho({\bm r_2},{\bm r_1})\!=\!\sum_\nu 
\phi_\nu^\dagger({\bm r_2})\phi_\nu({\bm r_1})$,
can be approximated as~\cite{Campi1978}
\begin{equation}
  \label{rmixed}
  \rho({\bm r_2},{\bm r_1})\approx \rho(z) \hat j_1(k_F s)\;,
\end{equation}
with $z\!=\!\textstyle{\frac12}|\bm r_1+\bm r_2|$, 
$s\!=\!|\bm r_1-\bm r_2|$,
and $k_F\!=\!(3\pi^2\rho)^{1/3}$, being $\rho$ the proton or neutron
density.
Additionally, $\hat j_1(x)\!\equiv\!3j_1(x)/x$.
In the case of a uniform sphere the mixed density in momentum space
gets expressed as~\cite{Arellano1990b}
\begin{equation}
  \label{mixed}
  \tilde\rho(q;P) = \tilde\rho(q)\times
  \frac{3}{4\pi\bar k_F^3}\,\Theta(\bar k_F-P)\;,
\end{equation}
with ${\bm q}$ and ${\bm P}$ the conjugate coordinates of
${\bm z}$ and ${\bm s}$, respectively.
The nonlocality of the mixed density is driven by its 
dependence on $P$ (denoted as $\mathcal{K}$ 
in Ref.~\cite{Burrows2018}),
limited by a bulk Fermi momentum $\bar k_F$.
As we shall see next, this feature becomes relevant for the
resulting nonlocality of the potential.

Let us now consider the Yukawa-type effective interaction
\begin{equation}
  \label{veff1}
  v_{\textrm{eff}}(r)=g_0\,Y(\mu r),
\end{equation}
with $Y(x)\!=\!e^{-x}/x$, where $g_0$ and $\mu$ stand for
strength and meson mass, respectively.
In momentum representation,
\begin{equation}
  \label{veff2}
\tilde v_{\textrm{eff}}({\bm p'},{\bm p})=
  \frac{1}{2\pi^2\mu}
  \frac{g_0}{({\bm p'}-{\bm p})^2+\mu^2}\;.
\end{equation}
Replacing this term in Eq.~\eqref{folding}, 
including its 
$\tilde v_{\textrm{eff}}({\bm p'},{-\bm p})$ exchange term,
we obtain for nucleon species $\alpha$ 
\begin{align}
  \label{asymm}
  U_\alpha(K,0)=& \frac{N_\alpha}{2\pi^2}\,
  \frac{3\,g_0}{4\pi \bm\bar k_F^3\mu}
  \int d{\bm P} \nonumber \\
  \times
  &
  \Theta(\bm\bar k_F-P)\, 
  \left [ \frac{1}{\mu^2}
    \pm \frac{1}{(\bm K-\bm P)^2+\mu^2}
    \right ]
  \;,
\end{align}
with $N_\alpha$ the number of protons or neutrons.
The $\pm$ signs depend on the spin-isospin parity 
of the channel involved, consistent with Pauli exclusion principle.
Considering singlet-even \textit{NN} states as leading contributions
to $\tilde U$, we select the plus sign.
The above integral can then be performed analytically, 
providing closed expressions for $\tilde H(K)$.
Additionally, a Taylor expansion of $U_\alpha$ in powers of
$K^2$ allows to extract $\beta$ as
function of the bulk Fermi momentum $\bm\bar k_F$ and mass $\mu$.
Details are given in \ref{toymodel}.
We use these analytic results to the case 
of the Michigan-three-Yukawa (M3Y) interaction~\cite{Bertsch1977}, where
\begin{equation}
  \label{m3y}
  v_{\textrm{eff}}(r)= \sum_{i=1}^{3} g_i\,Y(\mu_i r)\;,
\end{equation}
with $\mu_3\!=\!m_\pi$, the pion mass.
This third term accounts for the long range of the interaction, 
being dominant in the resulting range of nonlocality.

\begin{figure}[ht]
\includegraphics[width=0.95\linewidth] {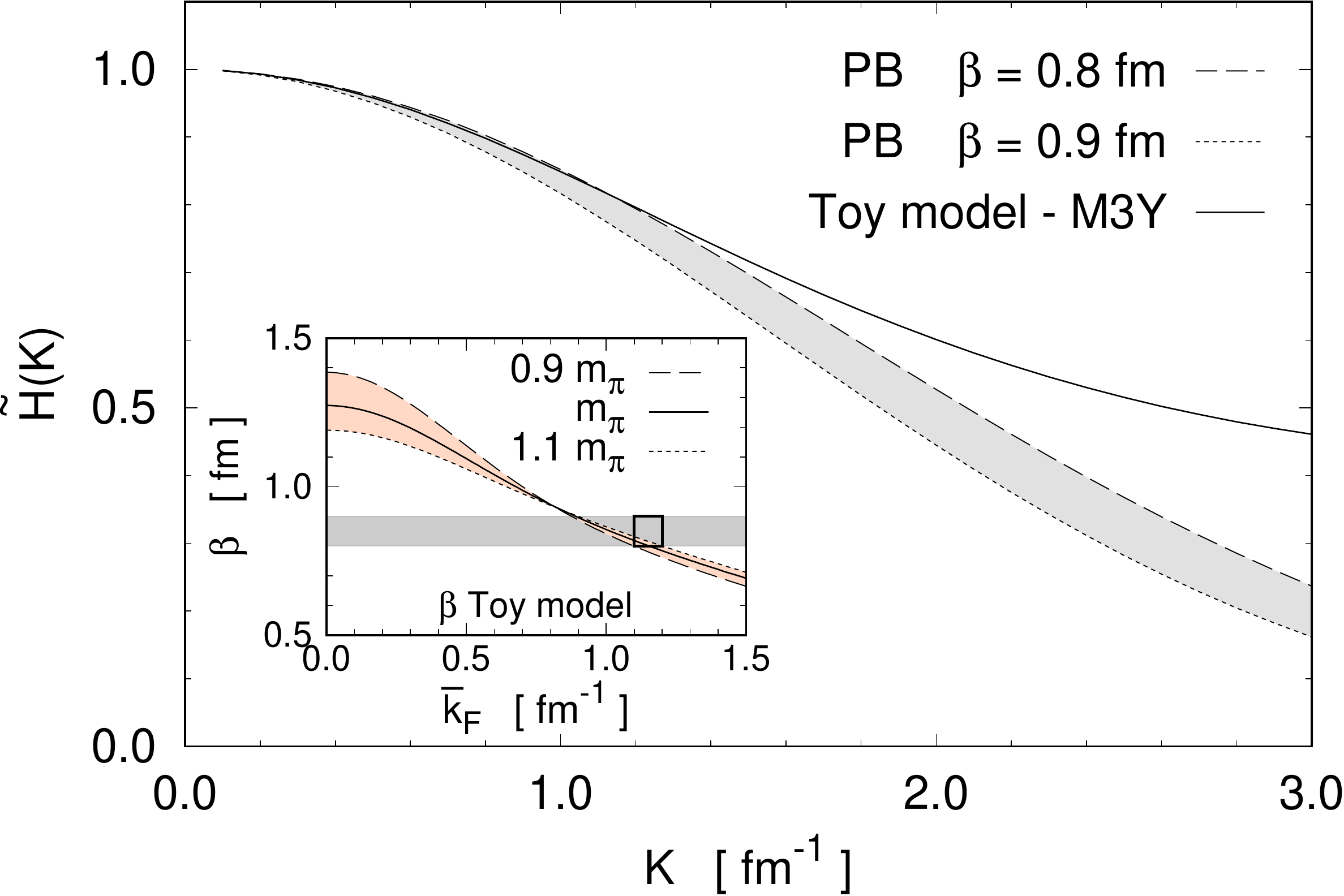}
\medskip
\caption{{\protect\small
\label{toy}
  Nonlocality form factor from toy model (solid curve).
  Dashed and dotted curves correspond to $\tilde H_{\!P\!B}$
  with $\beta\!=\!0.8$ and 0.9~fm, respectively.
  Inset shows $\beta$ as function of the bulk Fermi momentum
  $\bm\bar k_F$ in the toy model.
        }
        }
\end{figure}

In Fig.~\ref{toy} we plot the nonlocality form factor $\tilde H(K)$
as function of $K$ considering the M3Y interaction (solid curves).
Dashed and dotted curves represent Perey-Buck nonlocality
form factor $\tilde H_{\!PB}(K)$ with $\beta\!=\!0.8$ and 0.9~fm, 
respectively. 
We observe that, despite the simplicity of the toy model, 
it is reasonably consistent with PB for $K\!\lesssim\!1$~fm$^{-1}$.
The inset of Fig.~\ref{toy} shows the resulting $\beta$ as
function of the bulk Fermi momentum $\bm\bar k_F$.
To emulate changes in the long range part of the interaction
we include results with $\mu_3$ increased (dashed) and 
decreased (dotted) by 10\%.
The shaded strip denotes $0.8\!\leq\!\beta\!\leq\!0.9$~fm$^{-1}$,
with the box denoting $1.1\!\leq\!\bm\bar k_F\!\leq\!1.2$~fm$^{-1}$.
These bulk Fermi momenta are similar to those obtained
from density-weighted local Fermi momentum $k_z$ in
$^{40}$Ca and $^{208}$Pb, respectively.
It is interesting to observe, from this simple model, that
the range of nonlocality is mainly driven by the range of the
effective interaction together with the bulk Fermi momentum of the
nucleus.

\section{Conclusions}
Based on a momentum-space $g$-matrix approach for the optical model 
potential, we present the first quantitative account for Perey-Buck
range of nonlocality together with its implied form factor.
The nonlocality form factor exhibits a bell-shape, thought not
strictly Gaussian.
These findings result from an  analysis of the
structure of microscopic potentials based on chiral N3LO as well 
as AV18 \textit{NN} bare interaction models.
We find that,
when the potential is expressed as a function of momenta
${\bm q}$ and ${\bm K}$, 
its lowest order term in angular expansion leads to
the so called \textit{JvH} separable structure for
the central and spin-orbit parts of the potential.
In this factorization, $J$ corresponds to the volume integral
whereas $\tilde v$ and $\tilde H$ represent normalized
radial and nonlocality form factors, respectively.
We find that, while the radial form factor is usually
represented as Woods-Saxon form factor in phenomenological studies,
the nonlocality form factor exhibits close resemblance to that of
Perey and Buck~\cite{Perey1962}.

At nucleon energies below 65~MeV the range of nonlocality $\beta$ 
{for the central potential}
ends up in the range $0.86-\!0.89$~fm, in agreement with
phenomenological studies.
%
In the case of the spin-orbit component of the potential
the range of nonlocality is in the range 
$0.46-\!0.58$~fm, smaller than its central part. 
Beyond its ability to account for near-Gaussian nonlocality form 
factors and the accepted range of nonlocality $\beta$, 
the \textit{JvH} factorization offers a novel and well defined link
between theory and phenomenology.
The possibility to separate the coupling of the projectile with
target protons and neutrons offers an interesting route for exploring
nonlocality for isospin-asymmetric targets.
Thus, the \textit{JvH} factorization could be useful for assertive 
microscopic guidance 
to enhance phenomenological representations of nonlocality 
in \textit{NA} scattering phenomena. 

The microscopic optical model applied in this study,
summarized by Eq.~\eqref{folding}, has the advantage of 
requiring only radial densities. 
However, the more general model discussed in 
Refs.~\cite{Arellano2007a,Aguayo2008} 
allows for an explicit treatment of the
nonlocal one-body mixed density. 
Therefore, it would be of interest to assess the implications of
the full one-body mixed densities on the nonlocality form factor
of the optical potential.
%
%
%

\appendix
\numberwithin{equation}{section}

\section{Integrals for the toy model}
\label{toymodel}
To evaluate explicitly Eq. \eqref{asymm} for the toy model
using M3Y we pay attention to the direct and exchange terms
\begin{align}
  \label{fd}
  f_d(\bar{k}_F,\mu)&= \frac{3}{4\pi \bar{k}_F^3\mu}
  \int\! d\bm P\, \Theta(\bar{k}_F-P)\; \frac{1}{\mu^2}
  \\
  \label{fx}
  f_x(K,\bar{k}_F,\mu)&= \frac{3}{4\pi \bar{k}_F^3\mu}
  \int\! d\bm P\, 
  \frac{\Theta(\bar{k}_F-P)}{(\bm K-\bm P)^2 +\mu^2}
  \;.
\end{align}
For the direct term $f_d$ we obtain
\begin{equation}
  \label{fd2}
  f_d = \frac{1}{\mu^3}\;,
\end{equation}
whereas for the exchange term $f_x$ we get 
\begin{equation}
  \label{fx2}
  f_x = \frac{3}{2\bar{k}_F^3\mu} \int_{0}^{\bar{k}_F}\! P^2dP \!
  \int_{-1}^{1} \frac{du}{K^2+P^2-2KPu + \mu^2}\;.
\end{equation}
After angular integration we obtain
\begin{equation}
  \label{fx3}
  f_x=\frac{3}{4 k \bar{k}_F^3}
  \int_0^{\bar{k}_F/\mu} \! x\,
  \ln \!\left [
    \frac{(k+x)^2+1}{(k-x)^2+1}
    \right ] dx\;,
\end{equation}
where we have defined
\begin{equation}
  \label{k}
  k= \frac{K}{\mu}\;.
\end{equation}
The calculation of $f_x$ in Eq.~\eqref{fx3} can be performed analytically. 
If we denote
\begin{equation}
  \label{y}
  y=\frac{\bar k_F}{\mu}\;,
\end{equation}
after some algebraic work we obtain
\begin{align}
  f_x&= \frac{3}{2y^3\mu^3}
  \left \{
    y - \operatorname{Arctan}
    \left [\frac{2y}{1+k^2-y^2} \right] 
    \right .
    \nonumber \\
    &+
    \left .
    \frac{1+k^2-y^2}{4k}
    \ln\left[\frac{1+(k+y)^2}{1+(k-y)^2} \right] \right\}\,,
\end{align}
where the identity 
$\operatorname{Arctanh}(z)\!=\!\textstyle{\frac12}\ln[(1\!+\!z)/(1\!-\!z)]$,
was used.

Since leading contributions to the optical potential take place for
singlet-even \textit{NN} channels,
we select the plus sign for the exchange term in Eq.~\eqref{asymm}.
Thus, 
\begin{align}
  f_d+f_x &=\frac{1}{\mu^3}
    \left (
    1
    + \frac{3}{2y^3} 
    \left \{ 
    y - \operatorname{Arctan}
    \left [ 
    \frac{2y}{1+k^2-y^2}
    \right ] 
    \right . 
    \right . 
    \nonumber \\
    \label{fdx}
    &+ \
    \left . 
    \left . 
    \frac{1+y^2-k^2}{4k} \ln
    \left [ 
    \frac{1+(k+y)^2}{1+(k-y)^2} 
    \right ] 
    \right \} 
    \right ) \\
    &\equiv \frac{1}{\mu^3}\,F(y,k)\;.
\end{align}
To infer the low-$K$ behavior we expand $F(\mu,y,k)$ 
to second order in $k$,
\begin{equation}
  \label{f0f2}
  F(y,k)=F_0(y)-F_2(y) k^2 + {\cal O}(k^4)\;,
\end{equation}
where
\begin{subequations}
\begin{align}
  \label{f0}
  F_0(y) &= 
  1 + 
  \frac{3}{y^2} -
  \frac{3}{y^3} \operatorname{Arctan}(y)
  \;;\\ 
  \label{f2}
  F_2(y) &= \frac{1}{(1+y^2)^2}\;. 
\end{align}
\end{subequations}
In the context of the M3Y interaction,
the results in Eqs. \eqref{fdx}, \eqref{f0} and \eqref{f2}
have been used to obtain $U$ in Eq.~\eqref{asymm}, 
together with the implied nonlocality.




   \bibliographystyle{elsarticle-num} 

   \end{document}